\documentclass[11pt,twoside]{article}

\usepackage{asp2006_hotcool}
\usepackage{epsf}
\usepackage{graphicx}
\usepackage{lscape}

\begin{document}
\title{Tidal Flows from Asynchronous Rotation in Binaries}   
\author{Gloria Koenigsberger}
\affil{Instituto de Ciencias F\'{\i}sicas, Universidad Nacional Aut\'onoma de M\'exico, Cuernavaca, Mor., 62210, Mexico}
\author{Edmundo Moreno}
\affil{Instituto de Astronom\'{\i}a, Universidad Nacional Aut\'onoma de M\'exico, Mexico D.F, 04510, Mexico}
\author {David Harrington}
\affil{ Institute for Astronomy, University of Hawaii, 2680 Woodlawn Drive, Honolulu, HI, 96822, USA}

\begin{abstract} 
Asynchronous rotation in binary stars  produces non-radial oscillations that are known
to cause observable variability on  orbital timescales. The horizontal perturbations  of the 
surface velocity fields are  referred to as ``tidal flows". In this paper we illustrate the
manner in which tidal flows perturb the surface velocity field from that of uniform rotation,
using a one-layer stellar model for the calculations.  We justify the validity of this simplified 
model by the striking similarity between the photospheric absorption line-profiles it predicts  
and observational data of  the binary system $\alpha$ Virginis.  The velocity perturbations are  
used to compute the mechanical energy dissipation rates, $\dot{E}$,  due to the shearing flows   
for the case of  a massive (50$+$28 M$_\odot$) binary system having a moderately eccentric ($e=$0.3) 
orbit. The largest value of $\dot{E}$ around periastron phases is found on the hemisphere facing 
the companion.  However, at other orbital phases the maximum $\dot{E}$ may migrate towards  the poles.  
Assuming that $\dot{E}$ plays a role in the mass-loss characteristics of massive binary systems, 
this suggests that peculiar binaries such as HD~5980 and $\eta$ Carinae  may have a highly 
non-spherically symmetric mass-loss distribution which, in addition, is time-variable. 

\end{abstract}

\section*{Asynchronous rotation}

A system is in synchronous rotation when the  angular velocity of orbital motion $\Omega$ equals the
angular velocity of the star's rotation $\omega_0$.  In eccentric orbits, the degree of synchronicity varies with orbital
phase.  We use periastron passage as the reference point for defining the synchronicity
parameter $\beta_{per}=\omega_0/\Omega_{per}= 0.02 P v_{rot}(1-e)^{3/2} R_1^{-1}(1+e)^{-1/2}$,
where $e$ is the orbital eccentricity, the rotation velocity $v_{rot}$ is given in km/s,
the orbital period P is given in days, and the stellar equilibrium radius R$_1$ is given in solar units.
Non-radial oscillations are excited when at any orbital phase $\beta\neq$1.  The resulting 
surface velocity field consists of a pattern of perturbations superposed on the unperturbed 
stellar rotation field.  We refer to the horizontal components of the velocity perturbations, 
$\Delta V_\varphi$, as ``tidal flows".  A didactic manner of viewing the tidal flows is to
imagine that due to the perturbations, some portions of the surface may be rotating
slightly faster than the average rotation rate while other portions may be rotating slightly
slower.    The size and distribution  of these ``portions" depend on the modes in which 
the star is oscillating.

A huge body of research exists on the topic of tidal interactions, starting with
Darwin (1879, 1880). References and very interesting  discussions may be found in Eggleton 
et al. (1998) and Ferraz-Mello et al. (2008).  However, the focus is generally on 
the long-term evolution of the orbital parameters.  We in turn focus on  
the short-term effects, especially  those that may influence the observational properties 
of stars on orbital timescales.  

Our method consists of computing  the motion of a Lagrangian grid of surface elements distributed along
a series of parallels (i.e., rings with different polar angles) covering the  surface of a star of 
mass M$_1$ as it is perturbed by its companion of mass M$_2$.  The main stellar body below the 
perturbed layer is assumed to be in uniform rotation. The equations of motion that are solved for 
the set of surface elements include the gravitational fields of M$_1$ and M$_2$, the
Coriolis force, and the gas pressure. The motions of all surface elements  are coupled through the  viscous
stresses included in the equations of motion. The kinetic energy of the tides may be dissipated through
viscous shear, thus leading to an energy dissipation rate, $\dot{E}$.  The rate of energy dissipation per 
unit volume is given by the matrix product $\dot{E} = - {\bf P_{\eta}}:\nabla \mbox{\boldmath$v$}$,
where ${\bf P_{\eta}}$ is the viscous part of the stress tensor and {\boldmath $v$} is the velocity of
a volume element with respect to the center of the star (for details see Moreno \& Koenigsberger 
1999; Moreno et al. 2005; and  Toledano et al. 2007).

The benefits of our method are: 1) we  make no {\it a priori} assumption regarding the mathematical
formulation of the tidal flow structure since we derive the velocity field {\boldmath $v$} from
first principles; 2) the method is not limited to slow stellar rotation rates nor to small orbital
eccentricities; and 3) it is computationally inexpensive and contains only two free parameters,
the viscosity and the thickness of the surface layer.

  \begin{figure}
  \includegraphics[width=0.45\linewidth] {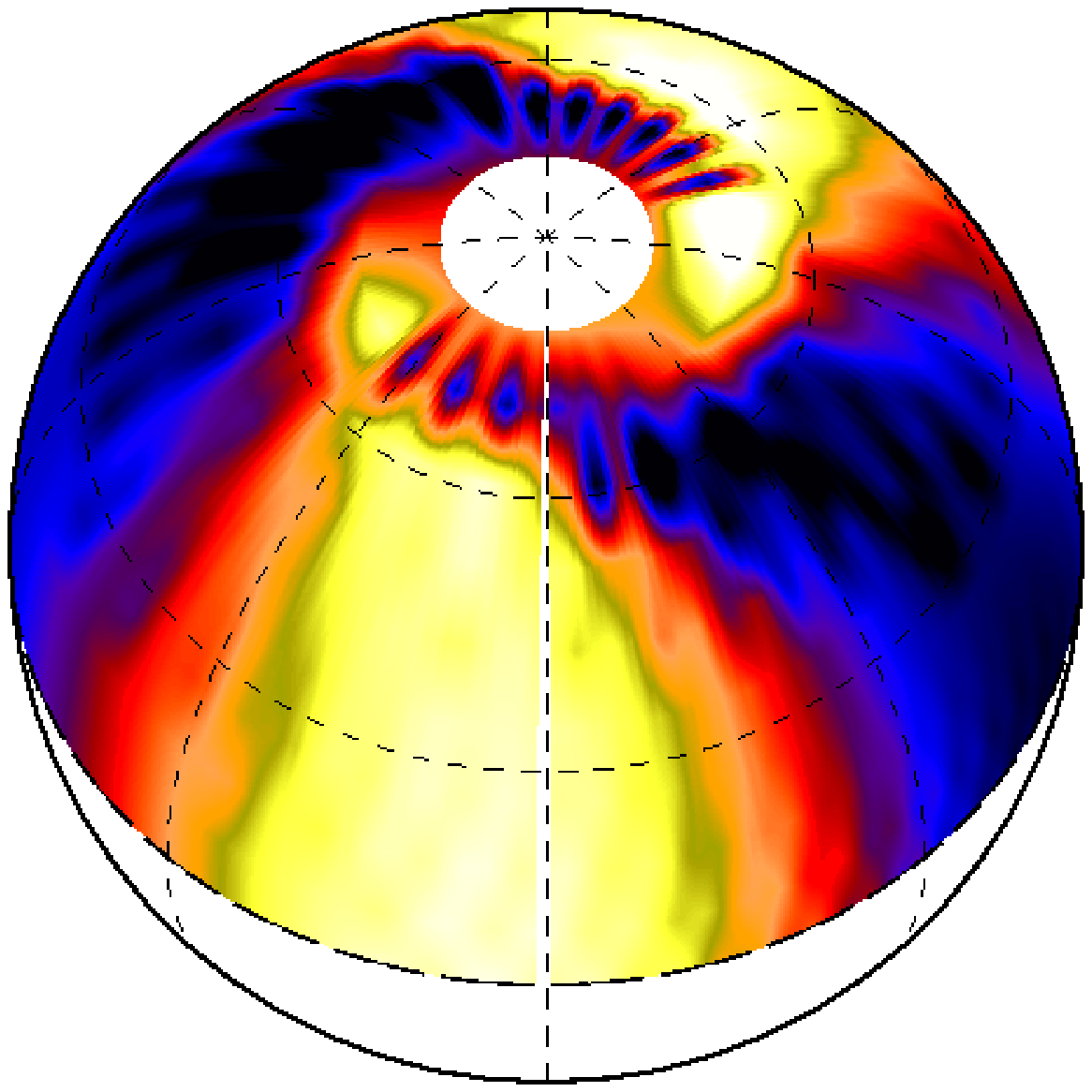}    
  \includegraphics[width=0.45\linewidth] {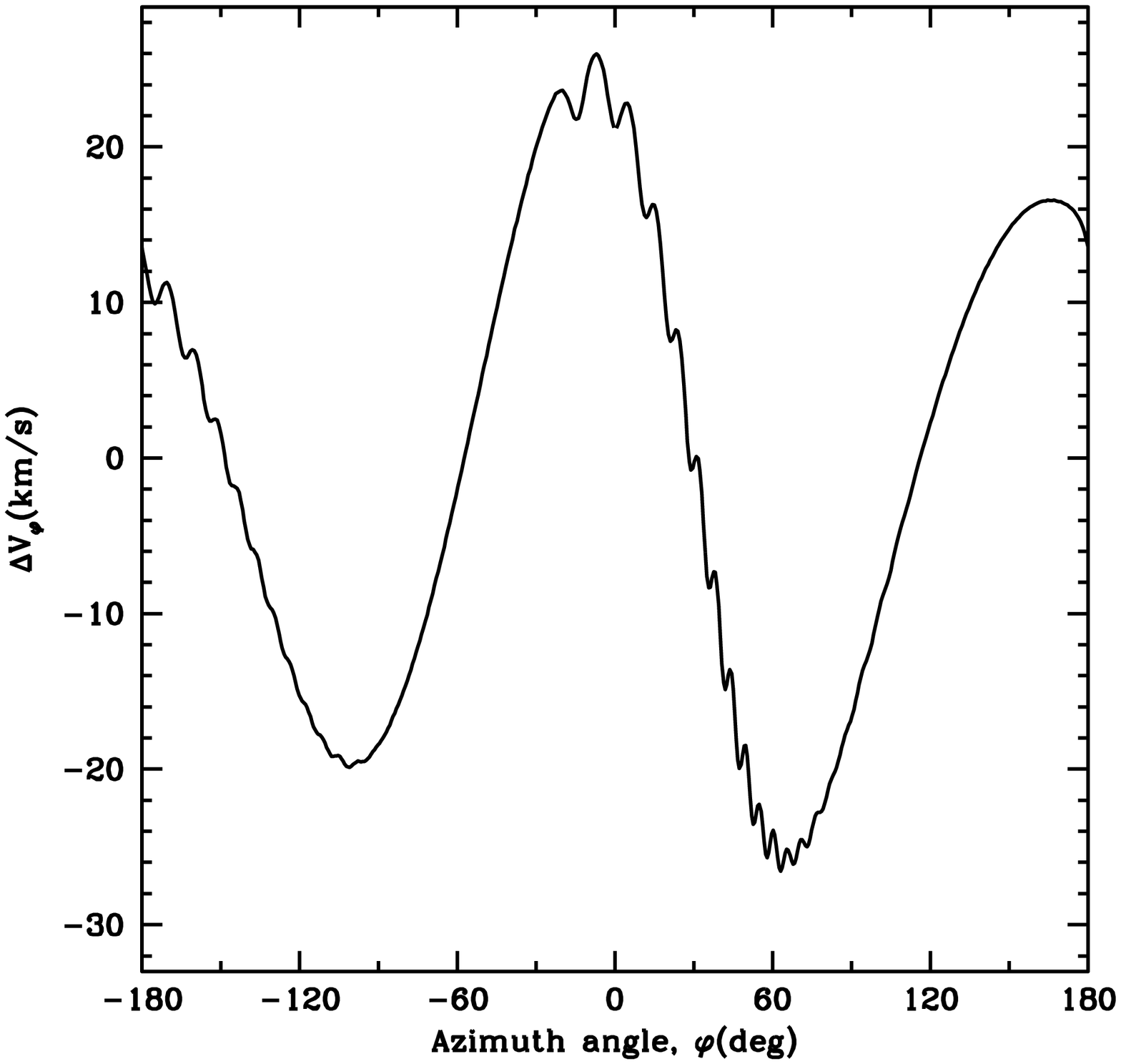}
  \caption{Left: Distribution of the azimuthal velocity perturbations,
   $\Delta V_\varphi$,  at periastron passage for the 50$+$28 M$_\odot$  
   model  discussed  in the text.  Color scale is such that white indicates maximum 
   perturbation  in the direction of rotation  while black indicates  maximum 
   perturbation in the opposite direction.   The sub-binary longitude is the vertical 
   line in the middle of the map; the rotation axis is tilted with respect to the plane 
   of the paper by 55$^\circ$. Right:  Plot of $\Delta V_\varphi$ for the equatorial 
   latitude, where positive values correspond to perturbations in the direction of rotation. }
   
  \end{figure}

  \begin{figure}
  \includegraphics[width=0.45\linewidth] {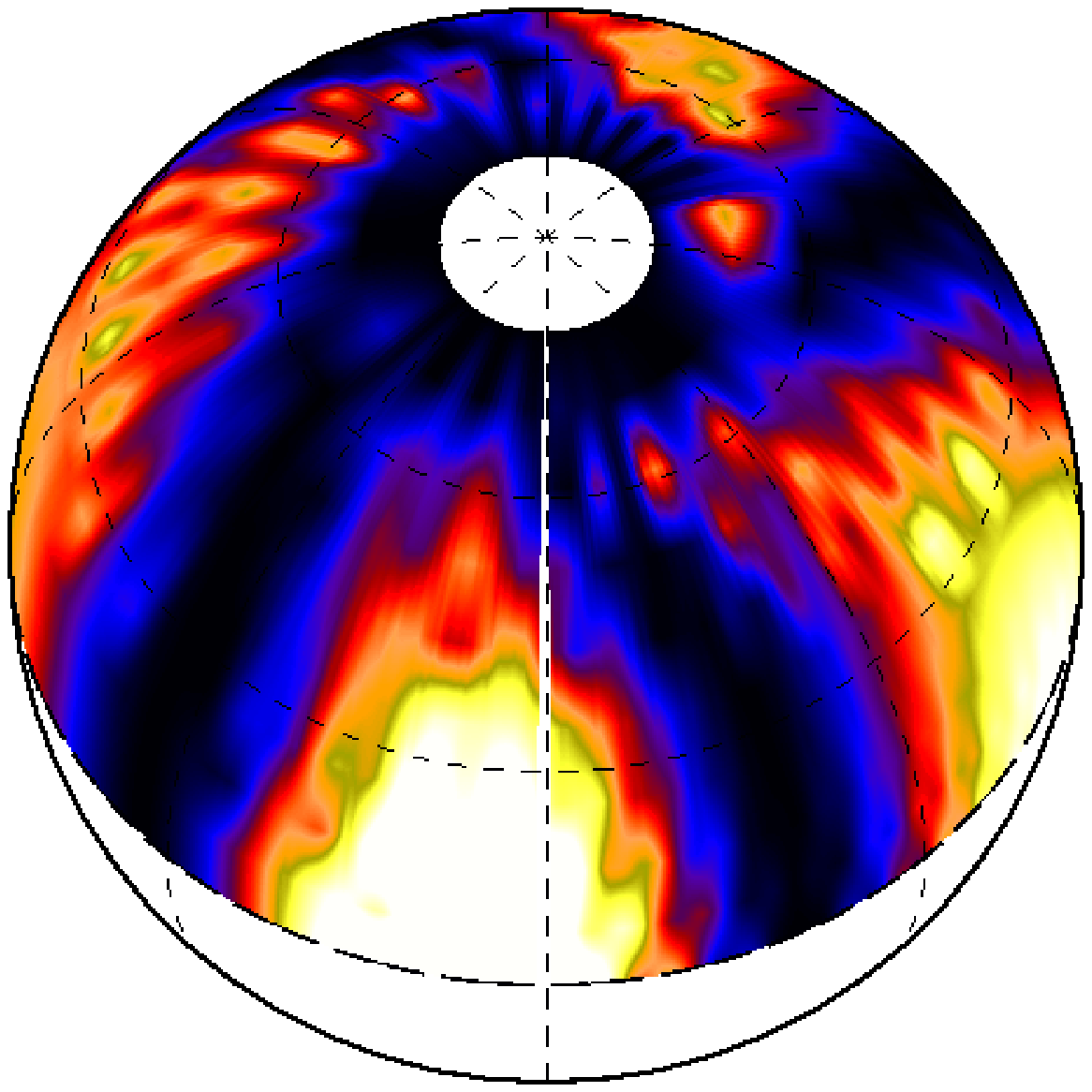}
  \includegraphics[width=0.45\linewidth] {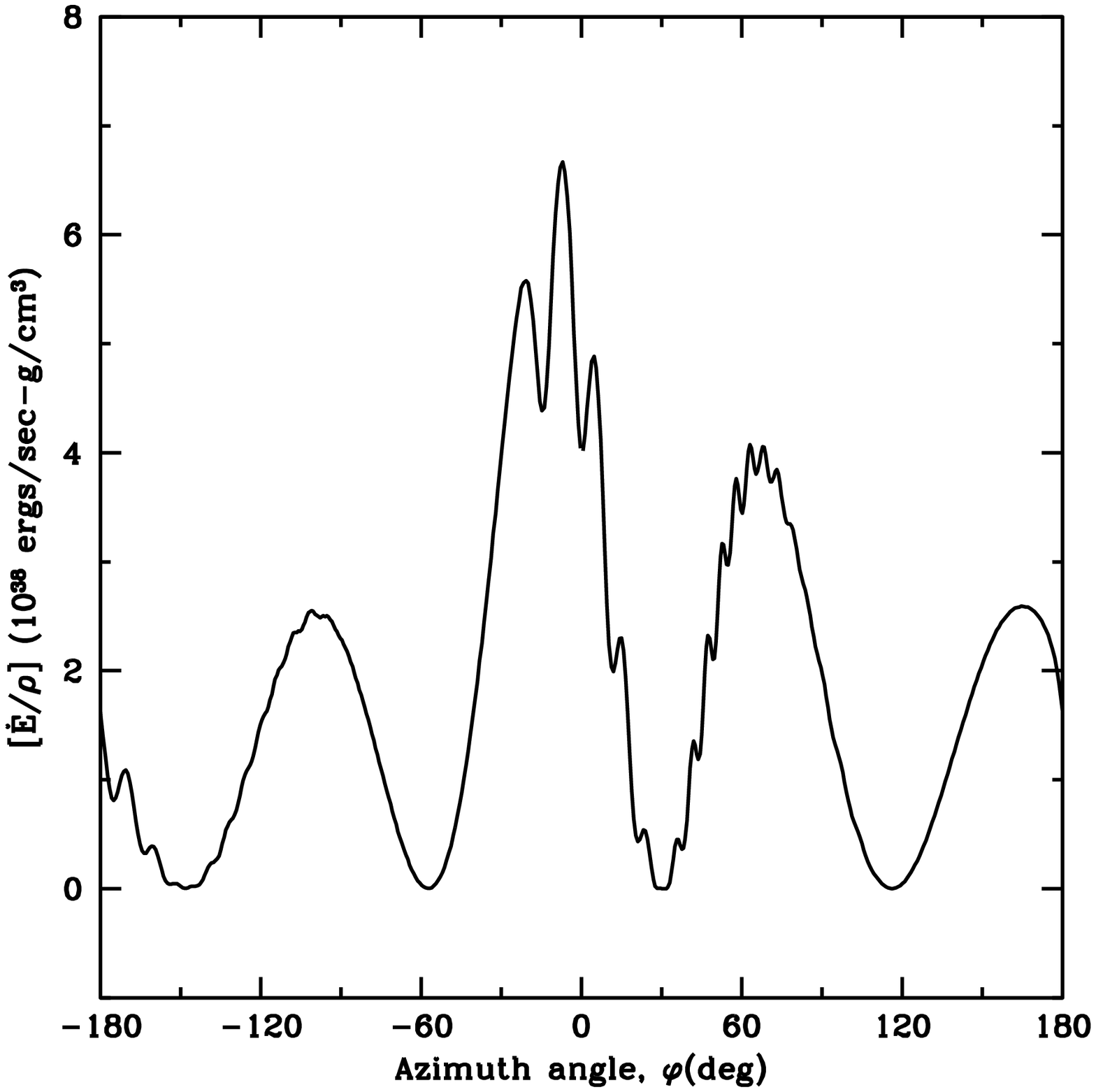}
  \caption{Same as previous figure except that here we illustrate the  
   distribution of the energy dissipation rate, $\dot{E}$, and the color scale
   such that white indicates maximum $\dot{E}$  and black indicates minimum values. 
    }

  \end{figure}

\section*{Surface Velocity Field and Energy Dissipation rates}

In Figure 1  we illustrate the characteristics of the perturbations of the 
azimuthal velocity field, $\Delta V_\varphi$ at periastron passage   for a 
M$_1+$M$_2=$50$+$28 M$_\odot$ binary system with P$=$19.3 d, e=0.3 and R$_1$=21 
R$_\odot$. Shown is the hemisphere containing the sub-binary longitude, $\varphi=$0. The
corresponding values of $\Delta V_\varphi$ on the equatorial latitude are plotted in
the right panel, illustrating the alternating positive and negative nature of this field.  
The largest  perturbation in the direction of stellar rotation occurs slightly west of the 
sub-binary longitude. The largest perturbation in the opposite direction occurs at a 
longitude of $\sim$60$^\circ$.  

The shear energy dissipation rate  $\dot{E}$ depends on the absolute value of the velocity 
perturbations, thus its azimuthal distribution has twice as many maxima and minima. A map
of $\dot{E}$ corresponding to the velocity field displayed in Figure 1 is in Figure 2 (left),
with the corresponding plot for the equatorial latitude (right).  Since $\dot{E}$ is
a function of $\Delta V_\varphi$, energy dissipation rates are large wherever the absolute 
value of the azimuthal velocity is large.

It is interesting to follow the changes over the orbital cycle in the $\dot{E}$ distribution.  A
selection of orbital phases is illustrated in Figure 3 for the same binary system  displayed in
Figures 1 and 2.  The maps are oriented so as to show the longitude range that would be visible
to an observer at the given orbital phases.  Shortly after periastron ($\phi=$0.10, bottom right),
the pattern remains similar to that shown in Figure 2 (periastron), although the actual values
of $\dot{E}$ are larger.  As the star approaches apastron, the values rapidly decrease and the 
pattern of maximum $\dot{E}$ breaks down into smaller zones and migrates towards the poles. After
apastron, the zones of large $\dot{E}$ begin to grow once again and migrate towards the equator.

  \begin{figure} [!h]
  \includegraphics[width=0.45\linewidth]{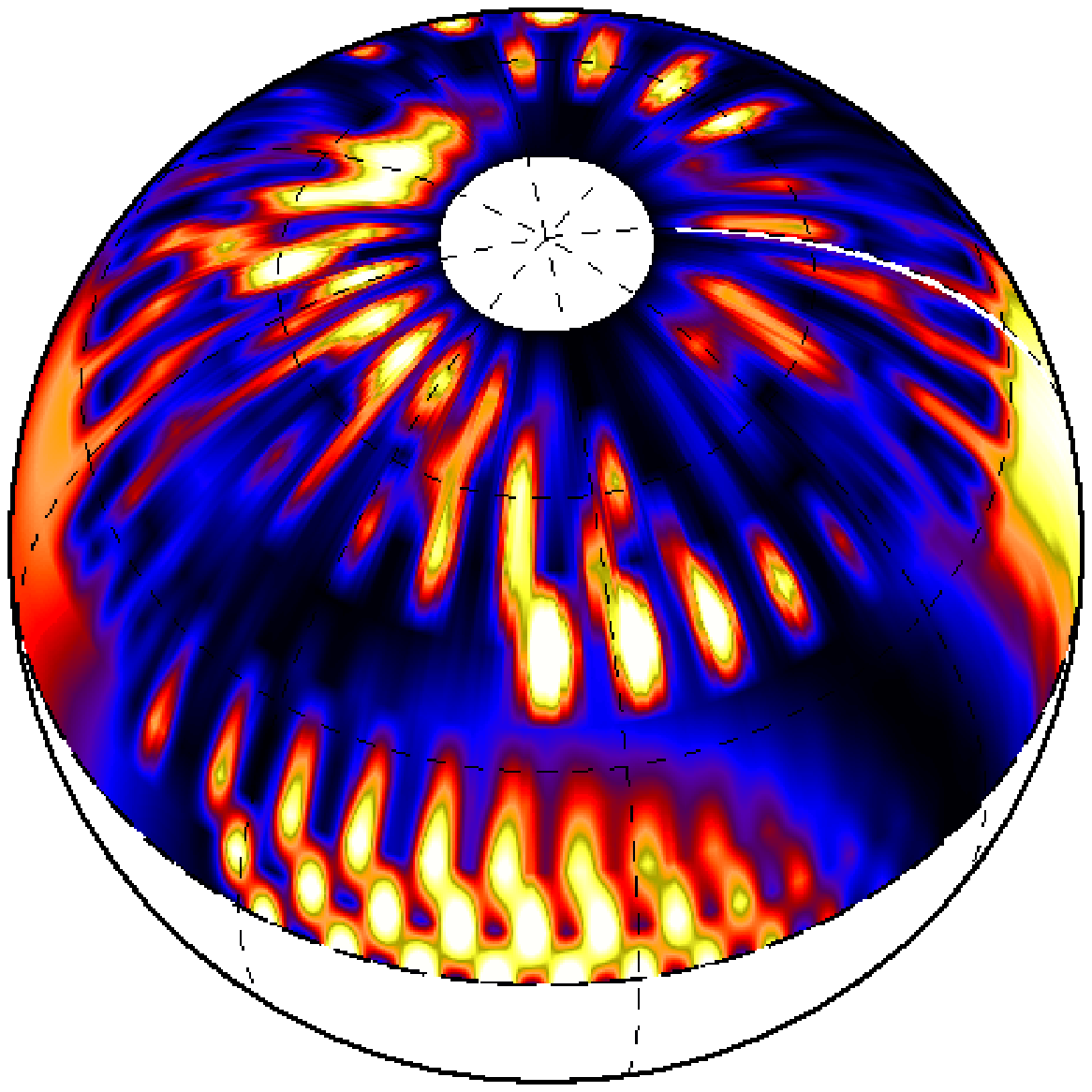} 
  \includegraphics[width=0.45\linewidth]{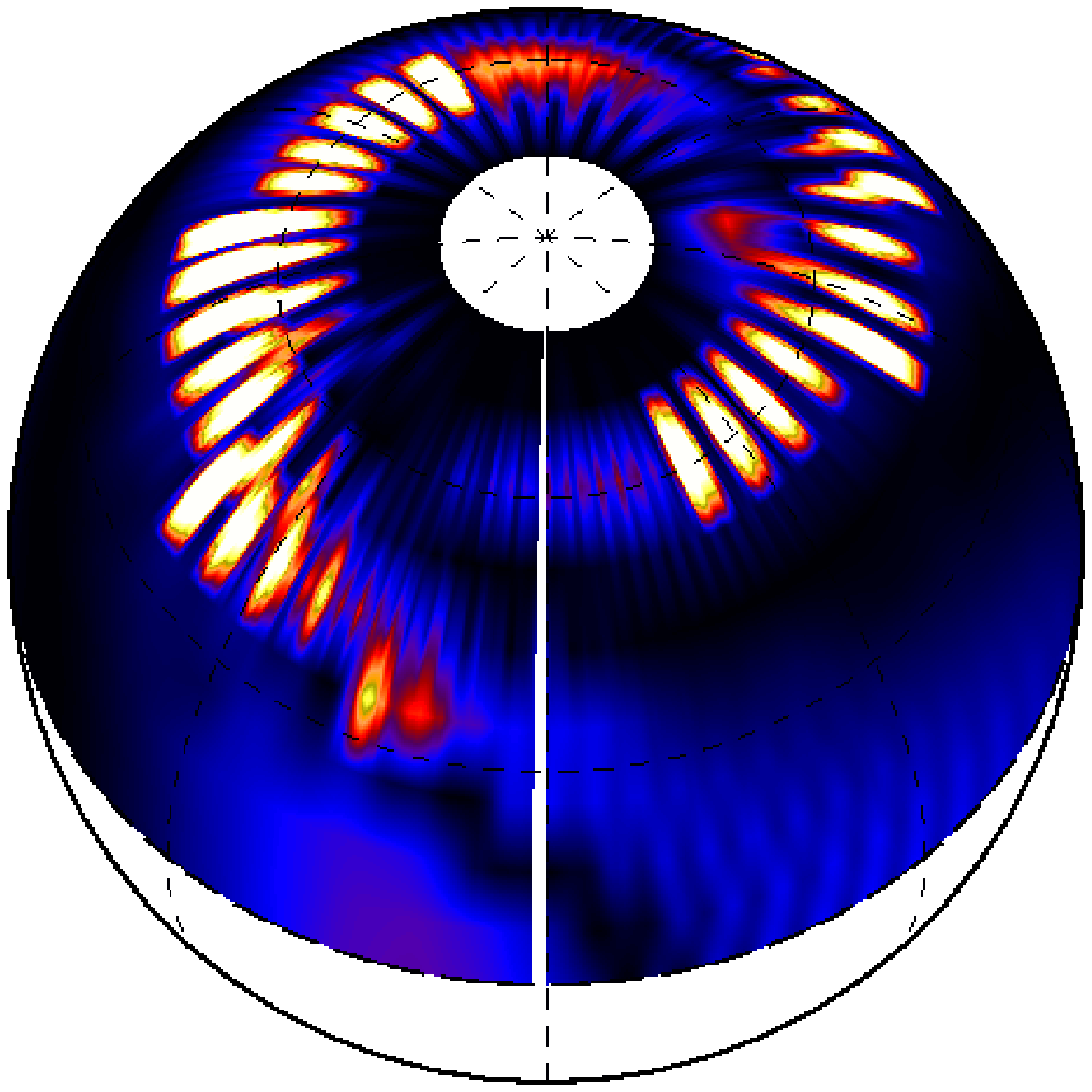}      
  \includegraphics[width=0.45\linewidth]{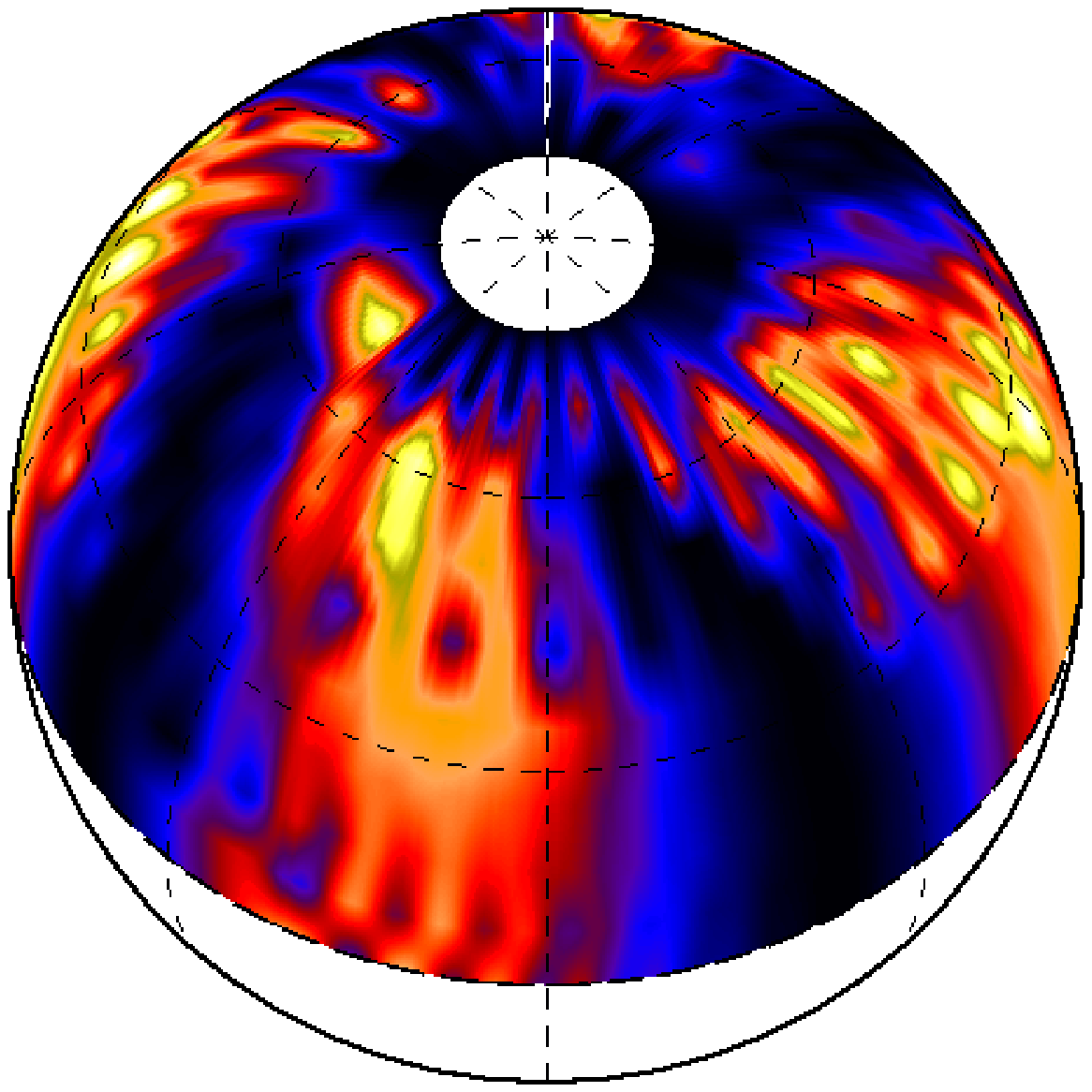} 
  \includegraphics[width=0.45\linewidth]{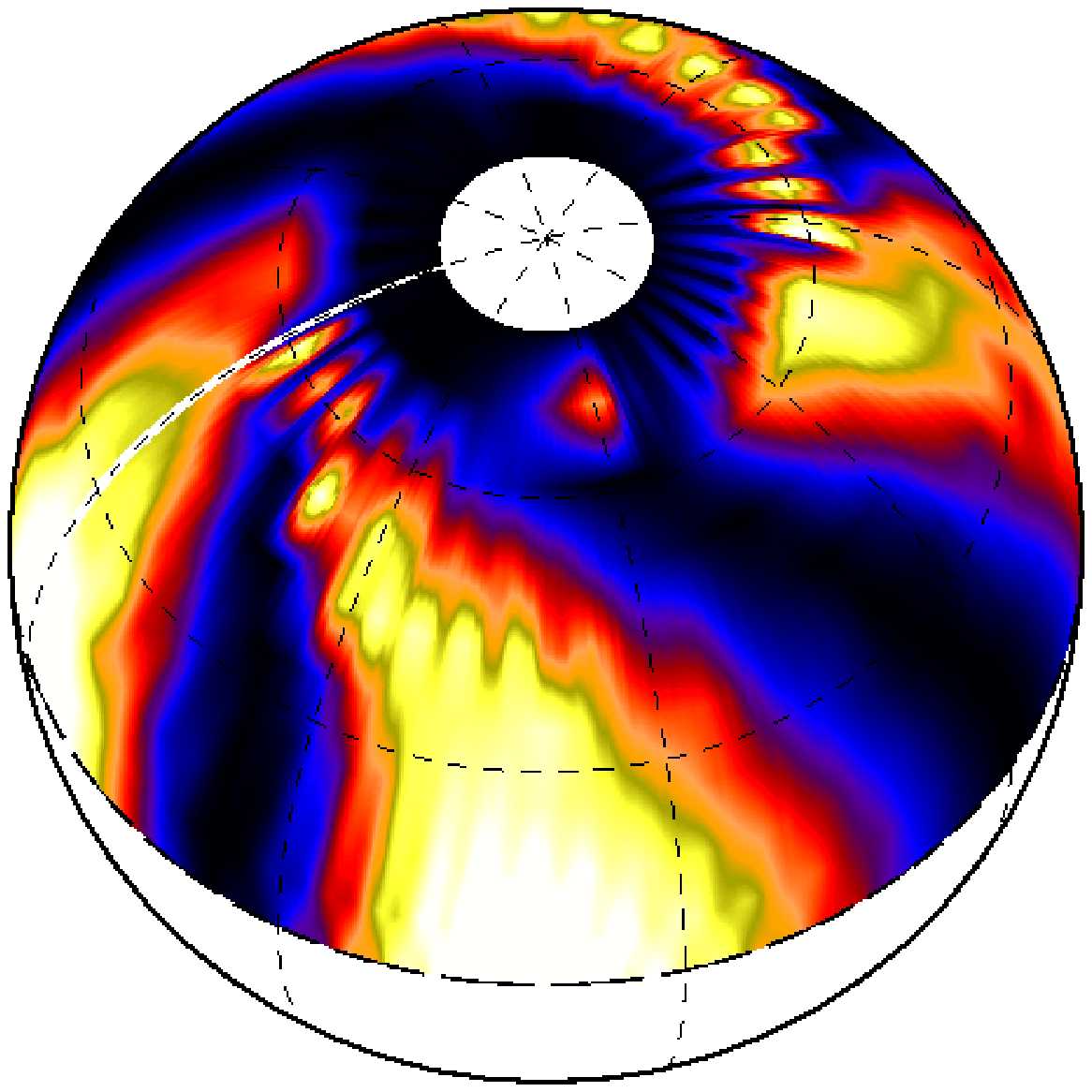} 
  \caption{Orbital phase-dependence of the $\dot{E}$ distribution  over the stellar surface
  for the same model illustrated in Figures 1 and 2 for  phases with respect to periastron 
  passage as follows: 0.10 (bottom right), 0.30 (top right)  and 0.80 (top left) and 0.95 (bottom left).  
  The maps are oriented so as to show the range in longitudes that would be visible to an external 
  observer and color coding is the same as in Figure 2.}
  \end{figure}

\section*{Photospheric line-profile variability}

A natural question that arises concerns the validity of the one-layer approximation
upon which our model is currently based.  One way to address this issue is by 
comparing the model results with observations of photospheric absorption line-profile 
variability.

Figure 4 illustrates the Si III 4552 \AA\ absorption line observed at four
different orbital phases in the B-type binary system $\alpha$ Virginis (Spica;
P=4.01 days, e=0.1).  The observations were performed  at the Canada France Hawaii  (CFHT) 3.6m
telescope with the ESPaDOns spectropolarimeter at a nominal spectral resolution of R=68,000
on 2008 March 20--28 (Harrington et al. 2009, in Preparation).  The   variability consists primarily
of the presence of   ``bumps" on the line profile that systematically change their 
location over time, as was first studied in detail by Smith (1985a, 1985b) is clearly seen.
Noteworthy also is the ``boxy" shape of the absorption at some orbital phases.
The dotted lines superposed on the observations correspond to the theoretical line
profiles computed with our model for the corresponding orbital phases. The model
reproduces both the general shape of the observational data as well as the number and
location of ``bumps".   Since this model is computed directly from first principles 
with only the single-layer approximation as an assumption and with very few free parameters,
the similarity between the computed and observed line profiles is striking.   The reason 
why the one-layer approximation works so well in this case is that the dominant effect 
producing the line profile variability arises from the external gravitational perturbation, 
rather than from the star's internal oscillation modes. 

  \begin{figure} [!h]
  \includegraphics[width=0.45\linewidth]{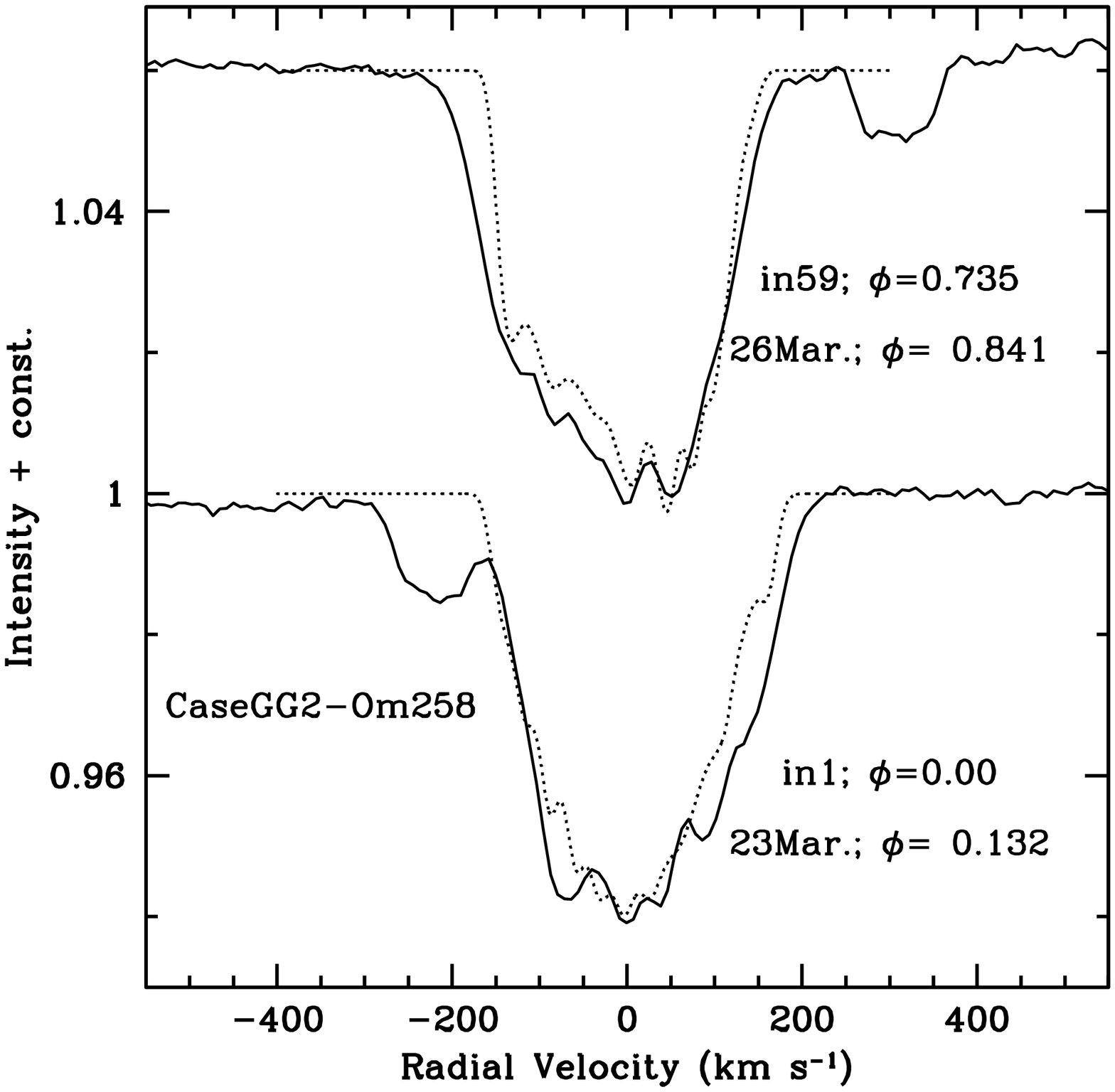}
  \includegraphics[width=0.45\linewidth]{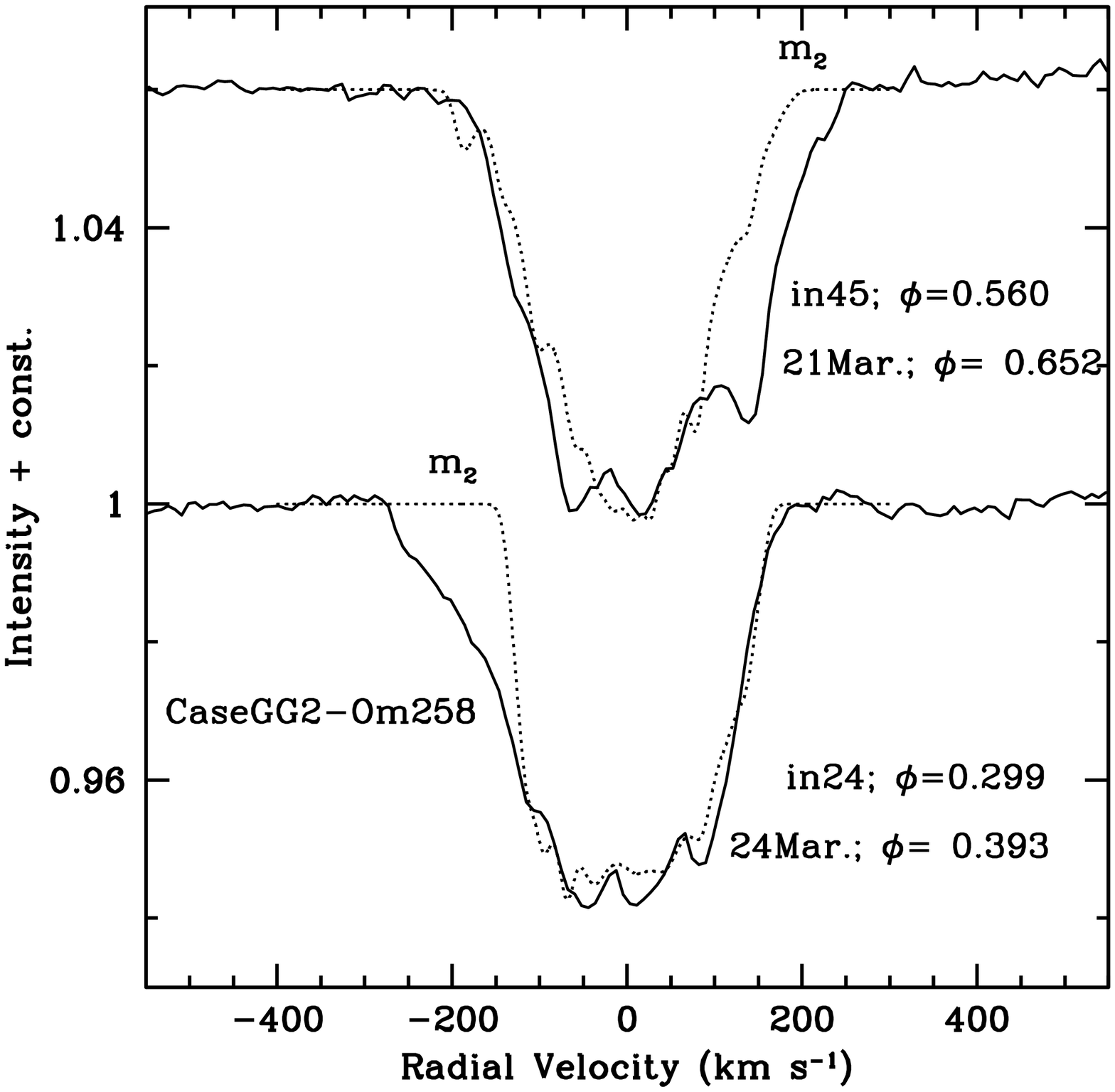}
  \caption{Comparison of observations  with theoretical (dots) line profiles for
   4 orbital phases in the binary systems $\alpha$ Virginis (Spica). Note
   that the one-layer model yields line profiles with a general shape that is 
   similar to that of the observations, and that the number and location of the sub-peaks also
   coincide.  The absorption line from the lower-mass companion is clearly visible in the
   left panel, but in the right panel it blends with the primary's absorption.  Its location
   is  indicated by the label ``$m_2$" and accounts for the larger discrepancy between the 
   observed and theoretical profiles in the right panel.}
  \end{figure}

\section*{Tidal flows and surface activity}

Although  synchronous rotation is generally assumed for the analysis of a large variety of 
binary systems, the uncertainties in the stellar radius, rotational velocity and orbital 
inclination  generally allow for significant departures from the assumed corrotating state.
Furthermore, no star in an eccentric orbit is in synchronous rotation.  Hence, the presence 
of tidal flows on the surface of binary stars is likely to be much more prevalent than 
generally assumed.

Tidal flows produce a surface velocity field having gradients
in both the azimuthal and polar directions.  In addition, due to the radial dependence 
of tidal forces, the velocity amplitudes on the surface are  larger than those 
in internal regions, hence also creating a rotation velocity gradient in the radial direction.  
All these gradients are strongly time-dependent.  We suggest that the strong and rapid variations
in these gradients may lead to a much larger degree of turbulence on the stellar surface than
is present in synchronized binaries or single stars.   It is also tempting to speculate that
these effects could  potentiate the appearance of localized magnetic fields (Ulmschneider \& Musielak 1998).

Tidal flows  lead to energy dissipation due to the associated viscous shear.
The additional energy in surface layers could conceivably alter the radial
temperature gradient as well as produce a non-uniform effective temperature distribution
over the stellar surface.  It is interesting to speculate on the possible effect that this additional
energy source could produce on stellar winds.   For example, since the tidal shear is not uniformly 
distributed over the stellar surface, the question arises as to whether $\dot{E}$ can contribute
towards enhancing the mass-loss rate.  Alternatively, localized magnetic fields or surface convective 
regions might tend to inhibit the radiatively-driven outflows at the base of the wind.  Hence, in
extreme cases where $\dot{E}$ is significant, our model would predict a highly non-spherically 
symmetric mass-loss rate.  In addition, the asymmetry would be greatest near periastron
passage, with a wind whose characteristics are dominated by the strong perturbations near
the equatorial latitudes while near aperiastron the effects of the perturbation might be
concentrated near the polar regions.

\acknowledgements 
We are grateful to Ben Brown for providing the idl scripts used to produce Figs. 1--3.
This investigation was partially supported by grants UNAM/PAPIIT 106708,  CONACYT 24936 and
NSF AST-0123390, the University of Hawaii and the AirForce Research Labs (AFRL).



\begin{thebibliography}{}
\bibitem[]{} Darwin, G. H., 1879 Philos. Trans. 170, 447
\bibitem[]{} Darwin, G. H., 1879 Philos. Trans. 171, 713
\bibitem[]{} Eggleton, P. P., Kiseleva, L. G. \& Hut, P. 1998, ApJ, 499, 853
\bibitem[]{} Ferraz-Mello, S., Rodr\'{\i}guez, A. \& Hussmann, H. 2008, CeMDA, 101, 171 
\bibitem[]{} Moreno, E. \& Koenigsberger, G. 1999, RMA\&A, 35, 157
\bibitem[]{} Smith, M. A. 1985a, ApJ, 297, 206
\bibitem[]{} Smith, M. A. 1985b, ApJ, 297,224
\bibitem[]{} Toledano, O., Moreno, E., Koenigsberger, G., Detmers, R., \&
             Langer, N. 2007, A \& A, 461, 1057
\bibitem[]{} Ulmschneider, P. \& Musielak, Z.E. 1998, A\&A, 338, 311

\end{thebibliography}
\end{document}